\documentclass[]{raa}            

\usepackage{graphicx,times}
\usepackage{natbib}
\usepackage{amssymb,amsmath}
\bibpunct{(}{)}{;}{a}{}{,}           
\usepackage{longtable}
\usepackage[pagebackref=true]{hyperref}

\def\cm2{cm$^{-2}$}

\def\nh3{NH$_3$}
\def\n2h{N$_2$H$^+$}

\def\13co{$^{13}$CO}
\def\c18o{C$^{18}$O}
\def\hc3n{HC$_3$N}
\def\h2{H$_2$}
\def\nh{n(H$_2$)}

\begin{document}

   \title{Tomography of the Ophiuchus Molecular Cloud with Velocity Features in C$_2$H $N=1-0$ spectra: A Pilot Study of Coherent Sub-structures}



   \volnopage{Vol.0 (200x) No.0, 000--000}      
   \setcounter{page}{1}          

   \author{Lei Qian*
      \inst{1,2,3,4},
   Zhichen Pan*
      \inst{1,2,3,4},
   Dongyue Jiang
      \inst{5},
   Zichen Huang
      \inst{6}}

   \institute{National Astronomical Observatories, Chinese Academy of Sciences,
             Beijing 100012, People's Republic of China; {\it lqian@nao.cas.cn; panzc@nao.cas.cn}\\
        \and
             Guizhou Radio Astronomical Observatory, Guizhou University,
             Guiyang 550025, People's Republic of China\\
	    \and
             College of Astronomy and Space Sciences, University of Chinese Academy of Sciences,
             Beijing, 100101, People's Republic of China\\
        \and
             Key Laboratory of Radio Astronomy, Chinese Academy of Sciences,
             Beijing 100101, People's Republic of China\\
        \and
             College of Physics, Guizhou University,
             Guiyang 550025, People's Republic of China\\
        \and
             Suzhou North America High School,
             No. 588 Lintian Road, Wuzhong District, Suzhou, 215104,
             People's Republic of China\\
   }

   \date{Received~~2009 month day; accepted~~2009~~month day}

\abstract{
The C$_2$H $N=1-0$ transition was used to investigate the possible line of sight sub-structures
from the dense and optically thick in $^{13}$CO $J=1-0$ regions in the Ophiuchus star forming molecular cloud.
With a 0.2 K or lower noise, multi-peak spectra were obtained and then used for identifying sub-structures.
There are clues, e.g., the core velocity dispersion remains unchanged with the increasing scale
that this cloud has a mild thickness in the line of sight direction and a large amount of overlapping CO cores,
as expected,
at least two coherent layers have been found.
The integrated intensity maps of these two layers are different in shape and morphology.
Inferred from the point velocity dispersion, one sub-structure with a thickness of $\sim 1$ pc was found,
while other substructures were more likely to be fragments.
\keywords{ISM: clouds; ISM: molecules; ISM: structure}
}

   \authorrunning{Qian}            
   \titlerunning{Asteroid}  

   \maketitle
%

%

\section{Introduction}

\label{sect:intro}

The Ophiuchus molecular cloud (or, $\rho$ Ophiuchi molecular cloud) is a low to medium mass star forming region with a distance of 137.3 pc \citep{2017ApJ...834..141O}.
This cloud consists of several Lynds sources, e.g., L1688, L1689, and L1709.
The dense regions in Ophiuchus have column densities of 1$\times$10$^{20}$ cm$^{-2}$ or even several orders higher,
judged from the extinction \citep{2017ApJ...834...91A}.
The star formation activities in the Ophiuchus molecular cloud were triggered more than once \citep[][and references therein]{2008hsf2.book..351W}.
The filamentary structures extended from either L1688 or L1689 were believed to be caused by those triggers, e.g., shocks from a supernova or an OB star association nearby \citep{2008hsf2.book..351W}.

From the study of the Core Velocity Dispersion \citep[CVD,][]{Qian2012, Qian2018, Qian2021},
the line-of-sight thickness of the Ophiuchus molecular cloud is estimated to be about 3.5 pc \citep{Qian2015}.
Besides the mild line-of-sight thickness,
it is also found that there are overlapping $^{13}$CO or dust continuum cores in dense region like L1688 and L1689 \citep{2000ApJ...545..327J}.
In addition, the $^{13}$CO $J=1-0$ lines in most part of the dense region of the Ophiuchus molecular cloud have double peaks.
It is thus very likely that there are multi-layered structures in the Ophiuchus molecular cloud.
The velocity coherency in the spectral map has been used to search for cores \citep{Qian2012} and filaments \citep{ 2014MNRAS.444.2507P} in molecular clouds,
with simultaneous fitting in velocity and space, i.e. in a p-p-v datacube.
For the search for features with larger scales, e.g. sheets,
it is attempted by fitting the spectrum at every point to decompose coherent features with different velocities.

The archival data of Ophiuchus include spectra lines mapping \citep[e.g., CO lines form COMPLETE survey,][]{2006AJ....131.2921R} and dust continuum mapping \citep{2000ApJ...545..327J}.
However, to catch possible sub-structures,
the data should have enough frequency resolution to separate these sub-structures (at least smaller than the line width, e.g., 1 km s$^{-1}$),
enough spatial resolution to identify cores (e.g., 1 arcminute or smaller),
and with relatively high sensitivity (e.g., noise level 0.2 K km s$^{-1}$ or lower).
In addition,
the transition should be with a wide range of volume densities (e.g., 10$^3$ to 10$^5$ cm$^{-3}$for optically thin) and
a wide range of evolutionary stages to cover the star forming history of the molecular cloud (e.g., from zero to 10 Myr).
Thus, we chose the archival Ophiuchus C$_2$H $N=1-0$ data from the 13.7-meter millimeter radio telescope at Delingha Observatory, Purple Mountain Observatory.
All these observations were done by the authors of this paper.

The C$_2$H gas is a chemical evolution tracer for carbon evolution \citep{2017ApJ...836..194P}.
In the early stages,
ultraviolet (UV) photons dissociate carbon monoxide, allowing the carbon to form other molecules.
Thus, the C$_2$H will remain in a relatively high abundance.
Along the evolution, the molecular cloud collapses and the extinction increases.
As the external UV photon can not go into the dense regions,
no more carbon was released from carbon monoxide.
Without the carbon supply, the abundances of C$_2$H in dense and high extinction regions will decrease.
In addition, within six hyperfine structures, the opacity of C$_2$H $N=1-0$ transition can be fitted.
These hyperfine structures separate at least several kilometers per second in the spectra,
making it easy to identify any possible multi-peak features.
So, C$_2$H exists in the low to medium density region and is an ideal tracer for substructure.

The content of this work is organized as follows.
The basic information on the data and data reduction procedures are presented in section 2.
Results and discussions can be found in Section 3.
The conclusions are presented in Section 4.

\section{Observation and Data Reduction}
\label{sect:data}

\subsection{Observation}

The Observations were performed with the 13.7-meter millimeter radio telescope in Delingha, Qinghai Station of Purple Mountain Observatory (PMODLH), Chinese Academy of Sciences.
The observations were done in the years 2012, 2013, and 2017.
The observation regions were extended from only L1688 and L1689 to the whole Ophiuchus region that has $^{13}$CO emission.
Details of the observations, including the source names used, the observing dates, the mapping areas, the center coordinates, the sampling time, and scan slew rate can be found in Table 1.

The 3$\times$3 multibeam array covering 85 to 115 GHz was used for mapping the Ophiuchus region.
Both the upper sideband and lower sideband were with 16384 channels across 1000 MHz bandwidth,
corresponding approximately 0.2 km s$^{-1}$ frequency resolution.
The N$_2$H$^+$ $J=1-0$ (at upper sideband) and C$_2$H $N=1-0$ (at lower sideband) transitions were observed simultaneously.
The observing frequency settings for all the observations are the same.
In this study, we used C$_2$H N=1-0 data only.

Within the declination of -24 to -25 degrees,
the elevation of the Ophiuchus in the observatory was lower than 30 degrees.
The observations were done when the mapping areas were close to the maximum elevation.
Depending on the weather, the system temperatures from different beams of the array in different observations varied around 170 K and always lower than 200 K.
Detailed information on system temperatures and the calibration of antenna temperatures can be found in our previous study \citep{2017ApJ...836..194P}.
The results including mapping areas, N$_2$H$^+$ $J=1-0$ and C$_2$H $N=1-0$ integrated intensity maps will be in our next paper (Jiang et al. in prep).

\begin{table}[!ht]
    \centering
    \begin{tabular}{ccccccc}
    \hline
        Source & Observation & Mapping & Center  & Center & Sampling  & scan rates   \\
        Names & dates & areas & RA(J2000) & Dec(J2000) &  time(min) &  (arcsec/s) \\ \hline
        L1688 & 2012-12 & 23034 & 16:26:16 & -24:20:54 & 23*2 & 50 \\
        L1688-3 & 2013-03 & 23660 & 16:24:16 & -24:20:54 & 60*2 & 75 \\
        L1688-4 & 2013-03 & 23107 & 16:26:16 & -24:50:54 & 60*2 & 75 \\
        L1688-5 & 2013-03 & 30286 & 16:28:16 & -24:50:54 & 60*2 & 75 \\
        L1688-6 & 2013-03 & 17522 & 16:24:16 & -24:50:54 & 60*2 & 75 \\
        L1688-7 & 2013-03 & 34501 & 16:30:16 & -24:30:00 & 60*2 & 75 \\
        L1688-8 & 2013-03 & 33428 & 16:32:16 & -24:30:00 & 60*2 & 75 \\
        L1688-9 & 2013-04 & 60575 & 16:30:16 & -25:00:00 & 60*2 & 75 \\
        RHOOPH\_1 & 2013-11 & 41853 & 16:26:16 & -24:20:39 & 60*2 & 50 \\
        RHOOPH\_2 & 2013-11 & 46312 & 16:28:16 & -24:20:39 & 60*2 & 50 \\
        RHOOPH\_3 & 2013-11 & 47579 & 16:26:16 & -24:50:39 & 60*2 & 50 \\
        RHOOPH\_4 & 2013-12 & 35735 & 16:28:16 & -24:50:39 & 60*2 & 50 \\
        RHOOPH\_5 & 2013-12 & 53592 & 16:32:00 & -24:29:45 & 60*2 & 50 \\
        RHOOPH\_6 & 2013-12 & 47583 & 16:32:00 & -24:59:45 & 60*2 & 50 \\
        G354+15.5 & 2017-05 & 17656 & 16:34:05 & -24:31:45 & 60*2 & 75 \\
        G354+16.5 & 2017-05 & 17716 & 16:30:11 & -24:01:45 & 60*2 & 75 \\
        G352.8+18 & 2017-08 & 11842 & 16:22:08 & -23:52:58 & 60*2 & 75 \\
        G353+15.2 & 2017-08 & 11705 & 16:32:07 & -25:28:03 & 60*2 & 75 \\
        G353+17.6 & 2017-08 & 17407 & 16:24:08 & -23:52:58 & 60*2 & 75 \\
        G353.4+17.3 & 2017-08 & 16324 & 16:26:08 & -23:52:58 & 60*2 & 75 \\
        G354.3+16.2 & 2017-08 & 14596 & 16:32:11 & -23:56:45 & 60*2 & 50 \\
        G354.5+15.2 & 2017-08 & 5933 & 16:36:05 & -24:28:00 & 60*2 & 50 \\
        G355+14.9 & 2017-08 & 17673 & 16:38:17 & -24:20:00 & 60*2 & 50 \\
        G355.4+14.6 & 2017-08 & 17209 & 16:40:29 & -24:10:00 & 60*2 & 75 \\
        G354.4+16.2 & 2017-09 & 2881 & 16:32:38 & -23:53:42 & 60*3 & 50 \\
        G355+16 & 2017-09 & 2935 & 16:34:18 & -23:37:06 & 60*3 & 50 \\
        G355.4+15.9 & 2017-09 & 3008 & 16:35:58 & -23:20:30 & 60*3 & 50 \\
        G355.8+14.3 & 2017-09 & 17257 & 16:42:41 & -24:05:00 & 60*2 & 50 \\
        G355.9+15.9 & 2017-09 & 3017 & 16:37:38 & -23:00:54 & 60*3 & 50 \\
        G356.4+15.8 & 2017-09 & 2956 & 16:39:18 & -22:42:00 & 60*3 & 50 \\
        G356.4+16.6 & 2017-09 & 6294 & 16:36:19 & -22:09:19 & 60*4 & 50 \\
        G356.9+15.6 & 2017-09 & 3015 & 16:40:58 & -22:25:00 & 60*3 & 50 \\
        G357.4+15.5 & 2017-09 & 2968 & 16:42:38 & -22:08:00 & 60*3 & 50 \\
        G357.8+15.4 & 2017-09 & 5767 & 16:44:09 & -21:50:00 & 60*3 & 50 \\
        G358.4+15.4 & 2017-09 & 6281 & 16:45:36 & -21:28:00 & 60*4 & 50 \\ \hline
    \end{tabular}
\caption{\label{tab:widgets}The regions with C$_2$H $N=1-0$ observation.}
\end{table}

\subsection{Data Reduction}

As only C$_2$H data were used for this study,
all the data reductions mentioned below are only for C$_2$H $N=1-0$ spectra.
More details, such as the study of processing N$_2$H$^+$ data, obtaining column densities, or fitting for the opacities will be in our next paper (Jiang et al. in prep.).

All the original spectra from one single observation were gridded in the servers in the Delingha Station.
The noise level, namely $\sigma$, was used as the weight for gridding.
The cell sizes were set to be 30 arcseconds.
After downloading all the data from the observations mentioned in Section 2.1,
the Gildas/CLASS\footnote{https://www.iram.fr/IRAMFR/GILDAS} software package was used for further processing.
All the spectra were combined to create the three dimensions (RA, DEC, and velocity) data cube.
The option {\it nogrid} was used for Gildas/CLASS routine {\sc xy\_map} so that the spectra will not be gridded again.
To make the estimate easier, the channel width was slightly smoothed to be 0.22 km s$^{-1}$.
The center frequency of the spectrum was set to be 87317.05 MHz,
being the rest frequency of the C$_2$H $N=1-0, J = \frac{3}{2}-\frac{1}{2}, F = 1-1$ transition.
The velocity range of each spectrum was -20  to 20 km s$^{-1}$,
only included one hyperfine structure of C$_2$H $N=1-0$ transition.
As a result, a data cube containing 93205 spectra was obtained,
covering a range of RA 16h25m58s to 16h27m31s and DEC -25$^\circ$56'31'' to -21$^\circ$25'31''.

Our study aims to confirm the existence of sub-structures.
Thus, the strongest hyperfine structure of C$_2$H $N=1-0$, $J = \frac{3}{2}-\frac{1}{2} F = 1-1$ (87317.05 MHz),
was used.
If there is only one structure of the molecular in the line of sight,
the spectra should only have one peak, as shown in Figure 1 (upper).
If the spectra have multipeak features (for example, Figure 1, middle and lower),
there should be at least two components of the molecular cloud in the line of sight.

\begin{figure}[h]
\centering
\includegraphics[width=0.9\linewidth]{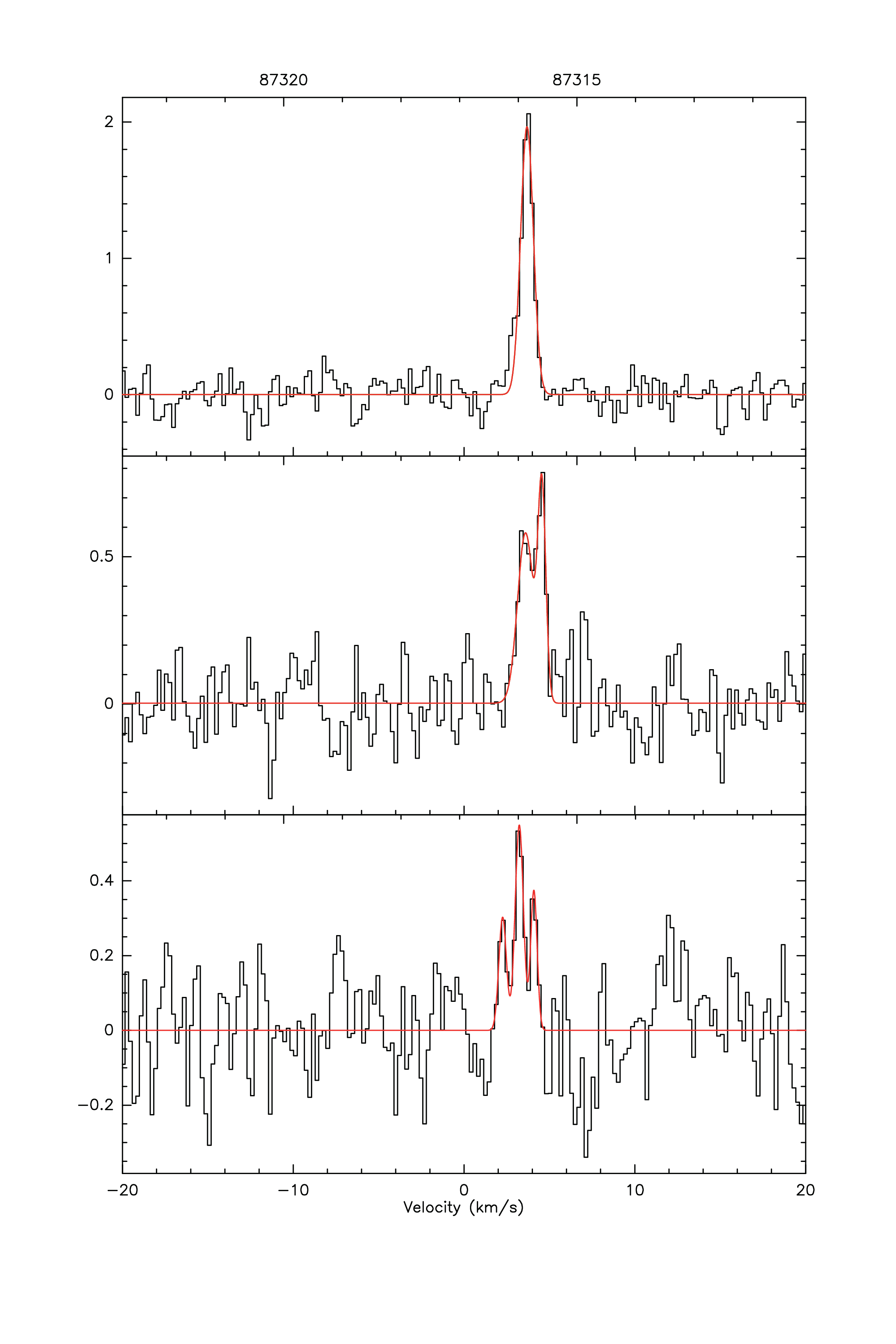}
\caption{Examples of the C$_2$H lined ($J = \frac{3}{2}-\frac{1}{2}$ $F = 1-1$, 87317.05~MHz) with one peak (upper), two peaks (middle), and multi peaks (lower). In the previous study, the root mean square value of the spectra from all the observing regions is 0.5 K or lower. With lower noise, the multi-peak feature is clearer.}
\label{multi}
\end{figure}

The integrated intensity is also affected by the noise.
For spectra with too low integrated intensity, the possible multi-peak feature may come from the noise.
Thus, the 0.5 K km s$^{-1}$ was used as the threshold.
As a result, 17914 spectra were selected from all the 93205.
Then, the Gaussian fitting was performed to obtain the parameters of each component in such spectra.

We manually fitted the peaks in all the 17914 spectra.
Firstly, the number of peaks in a spectrum was determined by the eye.
Secondly, the Gaussian fitting procedure from $class$ was used.
Thirdly, if the obvious peak(s) (e.g., 3 $\sigma$ from the peak-peak noise level) still existed in the residual,
the Gaussian fitting will iterate until a flat while noise spectrum is obtained.
The fitted centroid velocity, the line width, and the integrated intensity of each Gaussian peak were saved with the spectrum coordinates for identifying the sub-structures.

\section{Results and Discussions}
\label{sect:results}

\subsection{Detections of the ${\rm C}_2{\rm H}\ N=1-0$ Transition}

Previous studies reported no multi-peak features for the C$_2$H $N=1-0$ lines when the noise level is 0.2 to 0.5 K \citep{2017ApJ...836..194P}.
After observing the L1688 and L1689 regions for a longer time,
the noise level was reduced to 0.2 K or lower. Some spectra have more than one peak.
From the example spectra shown in the lower panel of Figure \ref{multi},
the spectrum has a peak intensity of 0.7 K and now clearly consists of two components.
The peak intensity of these two peaks are 0.6 and 0.8 K.
If the noise is 0.5 K, it will be merged to one peak.
Thus, it seems that the Ophiuchus cloud has different structures and can be more complex than being consists of these 'fibers'.

\subsection{Sub-Structures Identification}

\begin{figure}[h]
\centering
\includegraphics[width=0.5\linewidth]{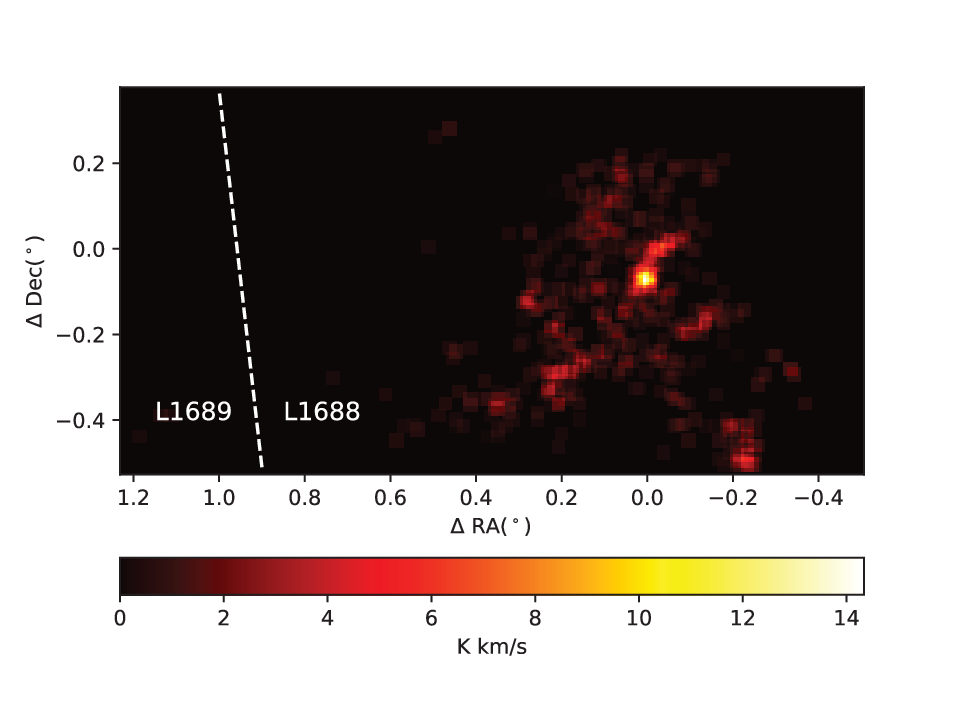}\includegraphics[width=0.5\linewidth]{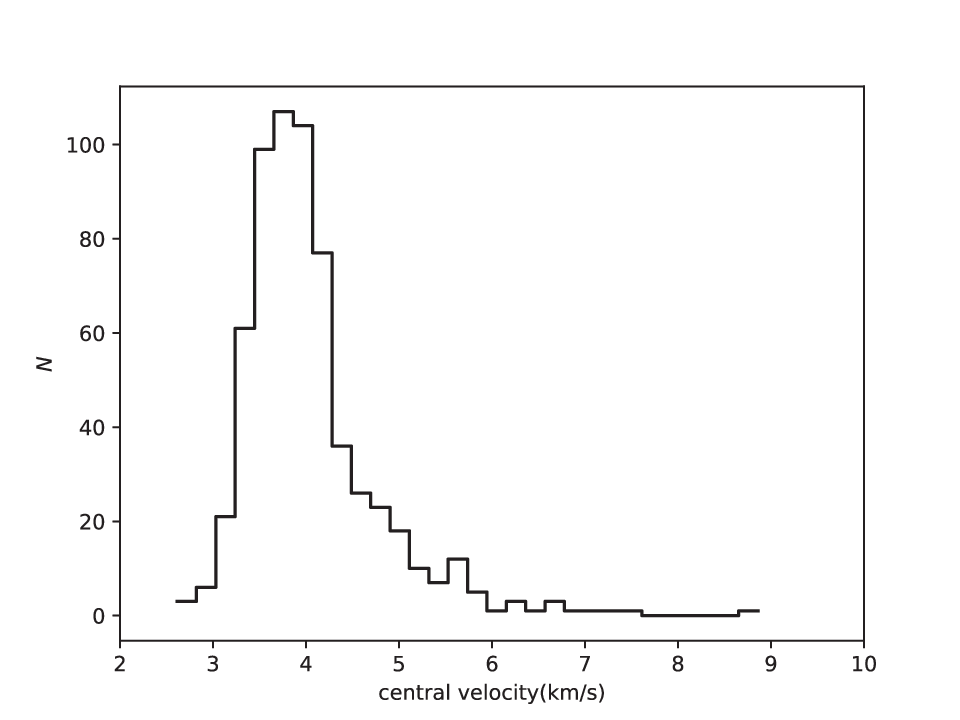}
\includegraphics[width=0.5\linewidth]{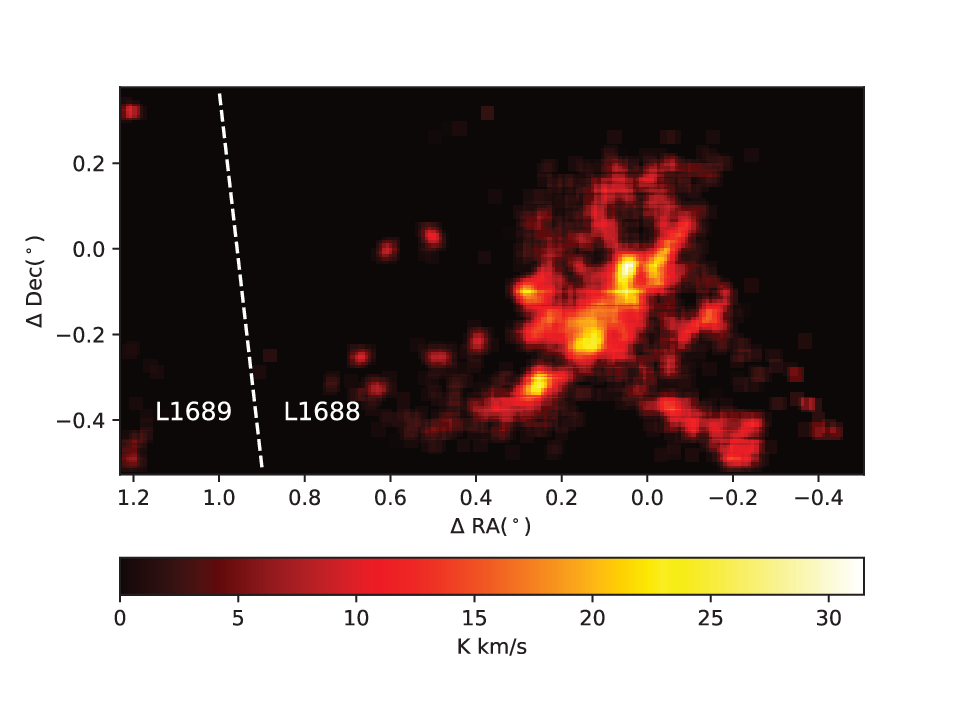}\includegraphics[width=0.5\linewidth]{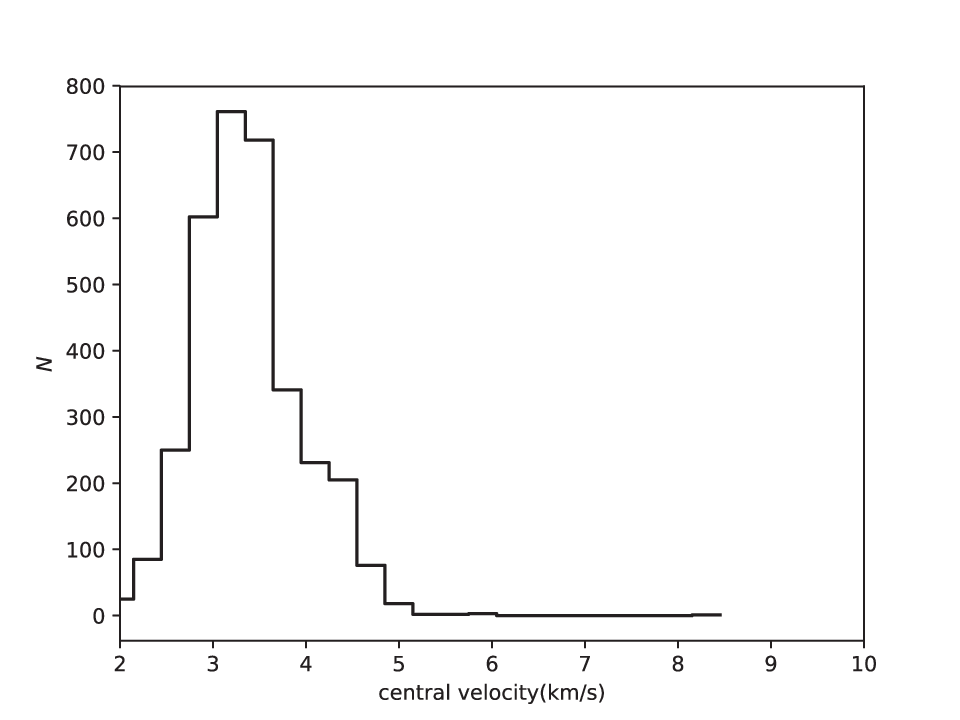}
\caption{The intensity map and the histogram of the velocities. The upper left panel shows a map of the integrated intensity of the C$_2$H component with a higher velocity ($\sim 3-6$ km/s) in the L1688 and L1689 region. The upper right plot shows the corresponding histogram of the velocities. The lower left panel shows a map of the integrated intensity of the C$_2$H component with a lower velocity ($\sim 2-5$ km/s). The lower right plot shows the corresponding histogram of the velocities. The $(0,0)$ point corresponds to RA=16:26:16, Dec=-24:20:54.}
\label{intensity_map}
\end{figure}

The possible sub-structures were identified by searching for coherent velocity components.
Currently, there is no known code that can do such an operation automatically.
Thus, we search these components by checking the data directly.
If a spectrum has more than one peak,
at most eight spectra around the position of the previous spectrum will be checked.
If any components in these spectra have similar centroid velocities and line width,
they were considered to come from the same sub-structure.
Iterating such operation along all the fitting results, we successfully identified at least two layers of sub-structures.
The intensities of the two Gaussian components at those points mentioned above are used to construct two intensity maps,
corresponding to a higher velocity ($\sim 3-6$ km/s, upper panel of Figure \ref{intensity_map})
and a lower velocity ($\sim 2-5$ km/s,lower panel of Figure \ref{intensity_map}), respectively.
The histogram of the centroid velocity of the two components are displayed accordingly in Figure \ref{intensity_map}.

It seems that sub-structures only appear in L1688,
because L1689 only appear in one panel of Figure \ref{intensity_map}.
Despite the compact core-like structures, there are extended structures identified in both intensity maps.
In the intensity map corresponding to the lower velocity component, the extended structures are more prominent than those in the map corresponding to the higher velocity component.
The sub-structures in the upper panel of Figure \ref{intensity_map} are not connected.
They may be different sub-structures.

\subsection{The Point (or Pixel) Velocity Dispersion and Comparison with Previous Results from Core Velocity Dispersion}

Previously, the structure of the Ophiuchus molecular cloud has been studied with the core velocity dispersion (CVD) technique in $^{13}$CO\citep{Qian2012}.
The CVD calculates the relationship between the projected distance of two cores and the velocity difference between them.
Theoretically, such a relationship can be expressed in a formula similar to Larson's Law \citep{1981MNRAS.194..809L}
\begin{equation}
    \sigma(v)\propto L^{\gamma}
\end{equation}
where the index $\gamma$ indicates the properties of the turbulence in the molecular cloud.
The $\gamma\sim 0.5$ is the currently accepted value \cite{1987ApJ...319..730S}.
If the cloud has a relatively large thickness,
the projected distance $L$ will deviate from the real distance.
In this case, CVD will not change with $L$. The power index will thus be close to zero \citep{Qian2015}.
It is found that the Ophiuchus molecular cloud has a moderate thickness of $\sim 3.5$ pc,
indicating possible complex structures\citep{Qian2015}.
However, it is hard to identify possible multi-layered structures with CVD,
because the $^{13}$CO spectra are optically thick. Not all the cores were identified.

To improve, we suggested calculating the Point (or Pixel) Velocity Dispersion (PVD) of the two C$_2$H gas components with a technique similar to CVD.
Here we did not fit Gaussian components to obtain the cores,
but calculate the velocity difference $\delta v$ and the distance $L$ of each pair of points.
The point pairs in the same distance bins are grouped to calculate the velocity dispersion
\begin{equation}
{\rm PVD}(L)\equiv \sqrt{\langle \delta v^2\rangle_{\tiny (L,L+\Delta L)} },
\end{equation}
where the angle brackets $\langle...\rangle$ denotes the average.
The distance of 137.3 pc \citep{2017ApJ...834..141O} was used to determine the distance between the two points in each point pair.

\begin{figure}[h]
\centering
\includegraphics[width=\linewidth]{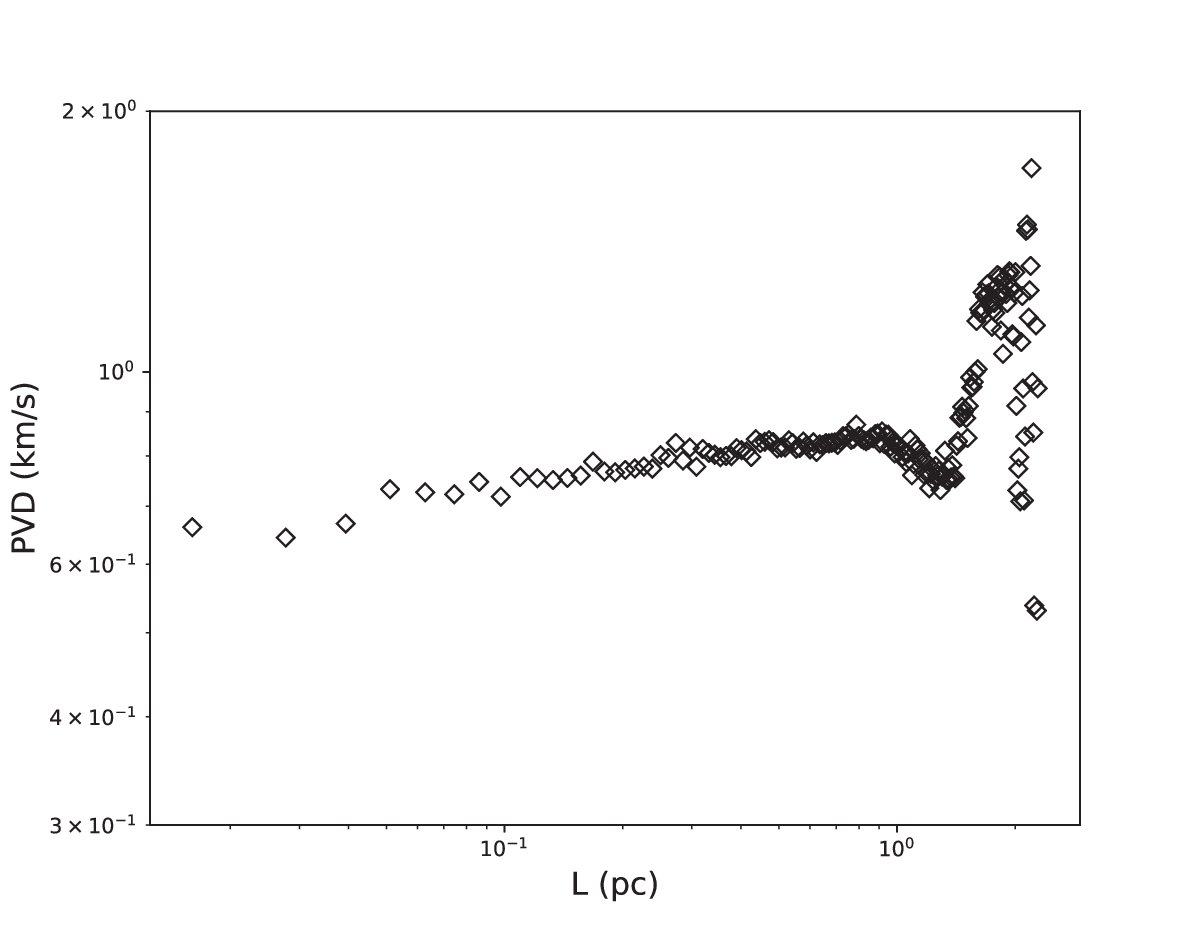}
\includegraphics[width=\linewidth]{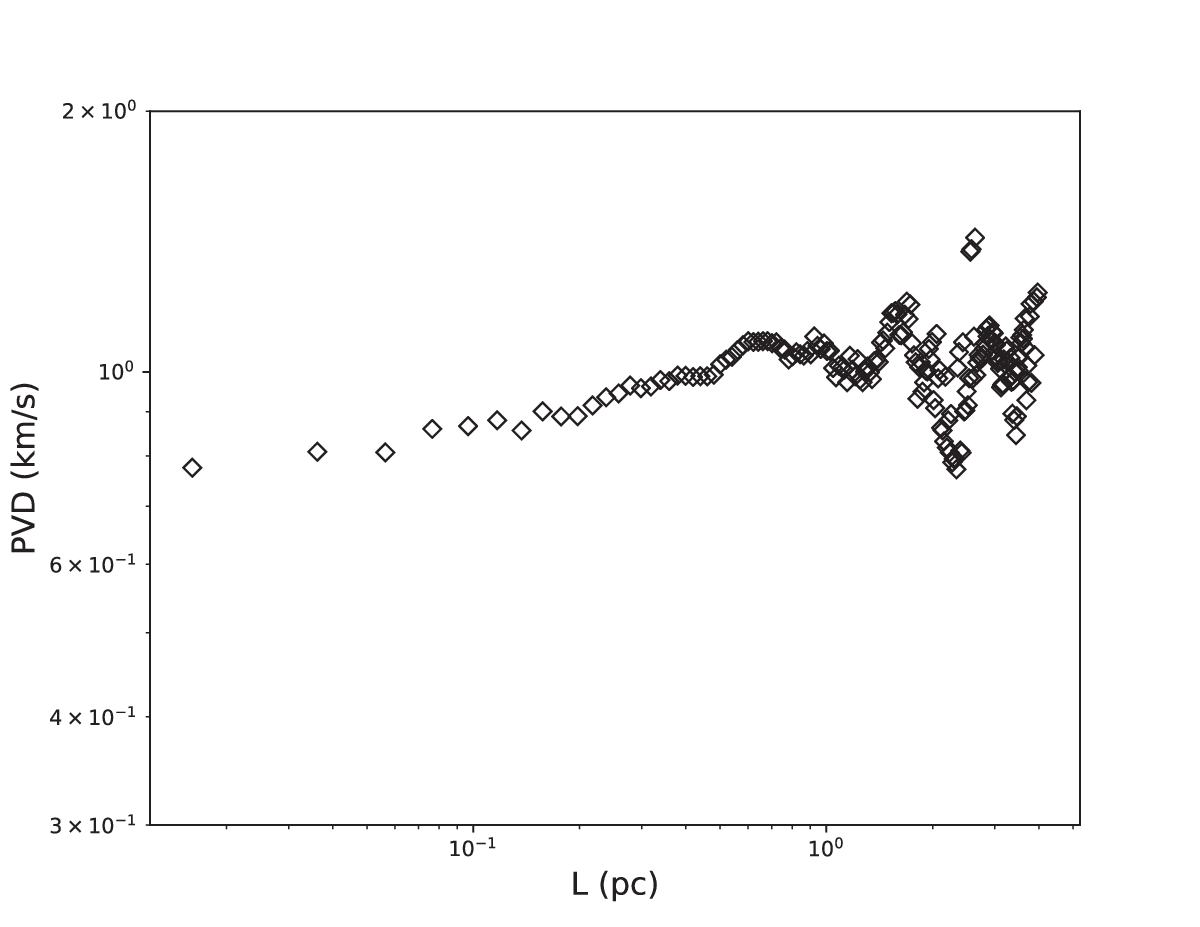}
\caption{The PVD v.s scale relation corresponding to the two maps in Figure \ref{intensity_map}. There is a break at $\sim 1$ pc in the PVD v.s scale relation of the C$_2$H gas with a higher velocity (upper panel). }
\label{pvd}
\end{figure}

The PVD plot of the two C$_2$H gas components can be found in Figure \ref{pvd}.
In the PVD plot of the C$_2$H gas components with higher velocity (upper panel),
there is a break at $\sim 1$ pc,
while there are no clear breaks in the PVD plot of the C$_2$H gas components with lower velocity (lower panel).
These results indicate that the C$_2$H gas components with higher velocity have a thickness of $\sim 1$ pc in the line of sight direction.
The thickness of another component is not determined from the PVD plot.
With different thicknesses, we suggest that the main part of Ophiuchus consists of at least two sub-structures.
One is similar to the one in the lower panel of Figure \ref{intensity_map},
while other sub-structures are more like small fragments separated from the main part of the Ophiuchus molecular cloud.

\section{Conclusion}
\label{sect:conclusion}

The conclusions of this work are as follows:

1, The C$_2$H $N=1-0$ observation of the whole Ophiuchus molecular cloud regions with $^{13}$CO $J=1-0$ emission was performed.
L1709 is the only new region with detectable C$_2$H $N=1-0$ emission.

2, In Ophiuchus molecular cloud, multi-peak features were detected from the spectra of C$_2$H $N=1-0, J = \frac{3}{2}-\frac{1}{2}, F = 1-1$ transition,
resulting in the discovery of at least two groups of sub-structures in the Ophiuchus molecular cloud.

3, L1688 is the only region with prominent sub-structures.
Inferred from the point (or pixel) velocity dispersion, the main part of L1688 should have a thickness of $\sim 1$ pc.
The other sub-strictures are more likely to be fragments.

As the star forming activities in the Ophiuchus molecular cloud were triggered several times,
future work will focus on the relationships between these sub-structures and the star formation triggers.

\vskip 1 cm
\noindent {\bf Authors Contributes}
\ Lei Qian analyzed the results, discovered the sub-structures, and provided the study results and discussions.
Zhichen Pan suggested this study and provided the data.
Dongyue Jiang controlled the data quality and filtered the spectra for fitting.
Zichen Huang worked on the spectra fittings and provided the fitting results.

\normalem
\begin{acknowledgements}

This work is supported by National SKA Program of China No. 2020SKA0120100,  National Nature Science Foundation of China (NSFC) under Grant No. 12003047, 11773041, U2031119, 12173052, and 12173053.
Both Lei Qian and Zhichen Pan were supported by the Youth Innovation Promotion Association of CAS (id.~2018075, Y2022027, and 2023064), and the CAS "Light of West China" Program. The observation was made
with the Delingha 13.7 m telescope of the Qinghai Station of
Purple Mountain Observatory (http://www.dlh.pmo.ac.cn), Chinese
Academy of Sciences. We appreciate all the staff members of
the observatory for their help during the observation. The
telescope and the millimeter wave radio astronomy database are
supported by the millimeter and submillimeter wave laboratory
of PMO.

\end{acknowledgements}


\begin{thebibliography}{9}
\providecommand{\natexlab}[1]{#1}
\providecommand{\selectlanguage}[1]{\relax}

\bibitem[Abrahams et al.(2017)]{2017ApJ...834...91A} Abrahams, R.~D., Teachey, A., \& Paglione, T.~A.~D.\ 2017, \apj, 834, 91. doi:10.3847/1538-4357/834/1/91

\bibitem[Galli et al.(2018)]{2018ApJ...859...33G} Galli, P.~A.~B., Loinard, L., Ortiz-L{\'e}on, G.~N., et al.\ 2018, \apj, 859, 33. doi:10.3847/1538-4357/aabf91

\bibitem[Goldsmith et al.(2008)]{2008ApJ...680..428G} Goldsmith, P.~F., Heyer, M., Narayanan, G., et al.\ 2008, \apj, 680, 428. doi:10.1086/587166

\bibitem[Johnstone et al.(2000)]{2000ApJ...545..327J} Johnstone, D., Wilson, C.~D., Moriarty-Schieven, G., et al.\ 2000, \apj, 545, 327. doi:10.1086/317790

\bibitem[Larson(1981)]{1981MNRAS.194..809L} Larson, R.~B.\ 1981, \mnras, 194, 809. doi:10.1093/mnras/194.4.809

\bibitem[Narayanan et al.(2008)]{2008ApJS..177..341N} Narayanan, G., Heyer, M.~H., Brunt, C., et al.\ 2008, \apjs, 177, 341. doi:10.1086/587786

\bibitem[Nutter et al.(2006)]{2006MNRAS.368.1833N} Nutter, D., Ward-Thompson, D., \& Andr{\'e}, P.\ 2006, \mnras, 368, 1833. doi:10.1111/j.1365-2966.2006.10249.x

\bibitem[\protect\citeauthoryear{Ortiz-Le{\'o}n et al.}{2017}]{2017ApJ...834..141O} Ortiz-Le{\'o}n G.~N., Loinard L., Kounkel M.~A., Dzib S.~A., Mioduszewski A.~J., Rodr{\'\i}guez L.~F., Torres R.~M., et al., 2017, ApJ, 834, 141. doi:10.3847/1538-4357/834/2/141

\bibitem[Pan et al.(2017)]{2017ApJ...836..194P} Pan, Z., Li, D., Chang, Q., et al.\ 2017, \apj, 836, 194. doi:10.3847/1538-4357/aa5c33

\bibitem[Panopoulou et al.(2014)]{2014MNRAS.444.2507P} Panopoulou, G.~V., Tassis, K., Goldsmith, P.~F., et al.\ 2014, \mnras, 444, 2507. doi:10.1093/mnras/stu1601

\bibitem[{{Qian} {et~al.}(2012){Qian}, {Li}, \& {Goldsmith}}]{Qian2012}
{Qian}, L., {Li}, D., \& {Goldsmith}, P.~F. 2012, \apj, 760, 147

\bibitem[{{Qian} {et al.}(2015){Qian}, {Li}, {Offner}, \& {Pan}}]{Qian2015}
{Qian}, L., {Li}, D., {Offner}, S., \& {Pan}, Z. C. 2015, ApJ, 811, 71

\bibitem[Qian et al.(2018)]{Qian2018} Qian, L., Li, D., Gao, Y., et al.\ 2018, \apj, 864, 116. doi:10.3847/1538-4357/aad780

\bibitem[Qian(2021)]{Qian2021} Qian, L.\ 2021, Acta Astronomica Sinica, 62, 7. doi:10.15940/j.cnki.0001-5245.2021.01.007


\bibitem[Ridge et al.(2006)]{2006AJ....131.2921R} Ridge, N.~A., Di Francesco, J., Kirk, H., et al.\ 2006, \aj, 131, 2921. doi:10.1086/503704

\bibitem[Solomon et al.(1987)]{1987ApJ...319..730S} Solomon, P.~M., Rivolo, A.~R., Barrett, J., et al.\ 1987, \apj, 319, 730. doi:10.1086/165493


\bibitem[Wilking et al.(2008)]{2008hsf2.book..351W} Wilking, B.~A., Gagn{\'e}, M., \& Allen, L.~E.\ 2008, Handbook of Star Forming Regions, Volume II, 351. doi:10.48550/arXiv.0811.0005



\end{thebibliography}


\end{document}